\documentclass[aps,prl,twocolumn]{revtex4}

\usepackage{graphicx,color}
\usepackage{hyperref}
\newcommand{\ba}{\begin{eqnarray}}
\newcommand{\ea}{\end{eqnarray}}
\newcommand{\be}{\begin{equation}}
\newcommand{\ee}{\end{equation}}
\newcommand{\bi}{\begin{itemize}}
\newcommand{\ei}{\end{itemize}}

\newcommand{\eps}{\varepsilon}

\begin{document}

\title{Probing confined phonon modes by transport through a nanowire double quantum dot}
\author{C. Weber}
\email[]{carsten.weber@teorfys.lu.se}
\author{A. Fuhrer}
\email[Now at: IBM Research-Zurich, R\"uschlikon, Switzerland ]
{afu@zurich.ibm.com}
\author{C. Fasth}
\author{G. Lindwall}
\email[Now at: Swerea KIMAB, 114 28 Stockholm, Sweden]{}
\author{L. Samuelson}
\author{A. Wacker}
\email[]{Andreas.Wacker@fysik.lu.se}
\affiliation{Nanometer Structure Consortium, 
Lund University, Box 118, 221 00 Lund, Sweden}

\date{February 15, 2010, published as Phys. Rev. Lett. 104, 036801 (2010)}

\begin{abstract}
Strong radial confinement in semiconductor nanowires leads to modified
electronic and phononic energy spectra. We analyze the current response to the
interplay between quantum confinement effects of the electron and phonon
systems in a gate-defined double quantum dot in a semiconductor nanowire. We
show that current spectroscopy of inelastic transitions between the two
quantum dots can be used as an experimental probe of the confined phonon
environment. The resulting discrete peak structure in the measurements is
explained by theoretical modeling of the confined phonon mode spectrum, where
the piezoelectric coupling is of crucial
importance.
\end{abstract}


\maketitle


The electronic confinement of semiconductor nanowires provides a
quantization of electron motion in the radial direction together with a 
quasi-one-dimensional dispersion in the axial 
wire direction. The resulting electronic 
properties have been intensively studied and suggest that nanowires are 
promising candidates for optical, electronic,
and thermoelectric applications \cite{AgarwalAP2006,
ThelanderMaterialsToday2006,BoukaiNature2008}. 

In analogy to the strong electronic confinement, the phonons are
similarly confined in nanowires, leading to a variety of modes with a
one-dimensional dispersion \cite{StroscioJAP1994Book2001}. This has been 
shown to yield distinct features in the excitonic absorption, such as
pronounced phonon replicas and a substantial broadening of the zero-phonon
line \cite{LindwallPRL2007,GallandPRL2008}. However, there have been only few
transport studies where features of electron-phonon coupling could be
identified. A recent example is the temperature
dependence of the resistance in Si nanowires which follows 
a power law characteristic for phonon
scattering \cite{VauretteAPL2008}. Going beyond such
averaged properties, 
we demonstrate in this letter that the
{\em individual phonon modes} resulting from the 
radial confinement of a nanowire
can be probed in a transport experiment. This allows for a new kind of phonon
spectroscopy which provides information on the energetic location of the phonon
modes as well as their interaction strength with the electrons. Such
information is crucial for any modeling of electronic or thermoelectric 
applications.

\begin{figure}
\includegraphics[width=0.95\linewidth]{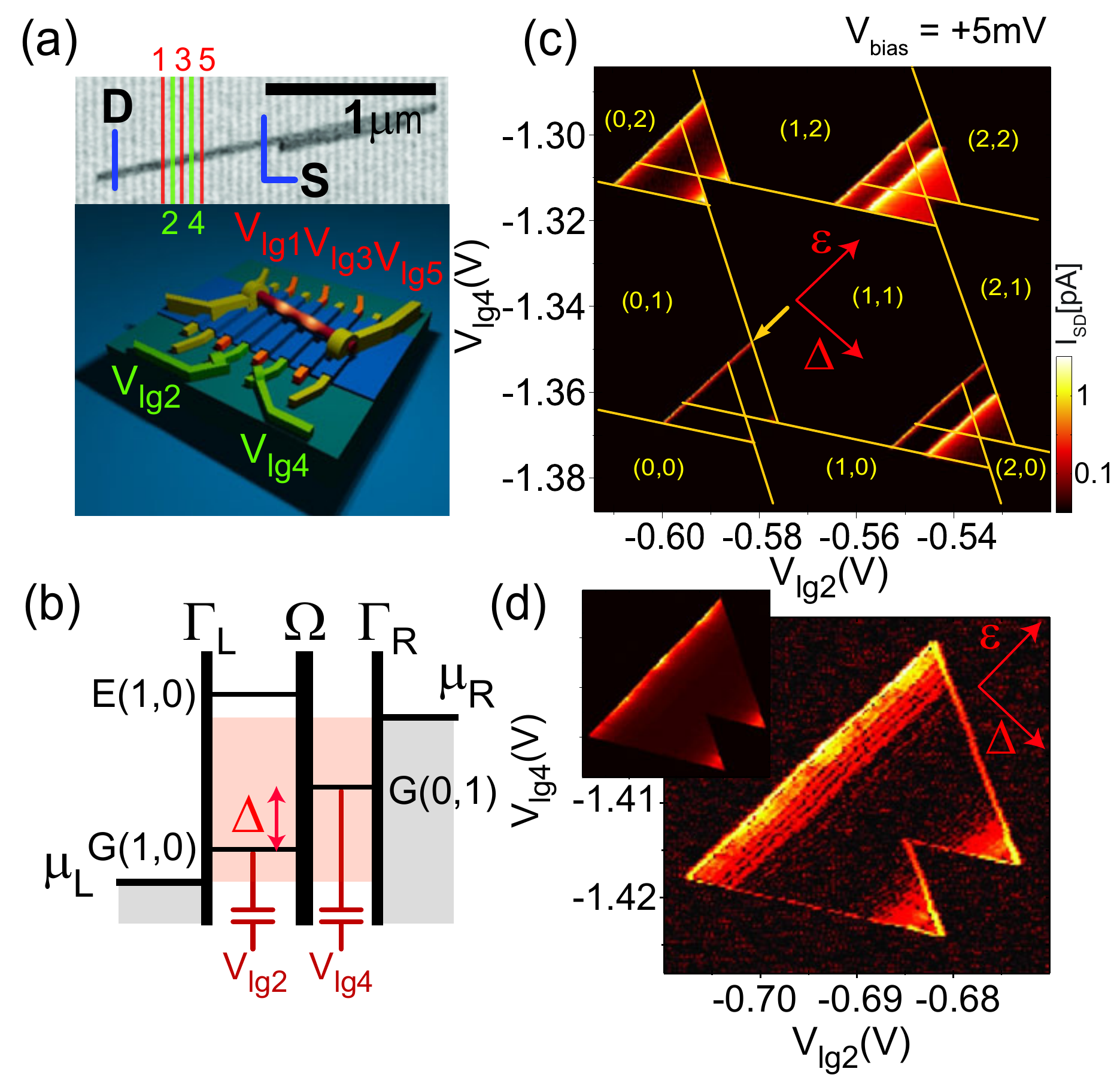}
\caption{\label{FigExpStruct} (Color online) (a) Scanning electron microscope image of the
measured nanowire lying on the gate electrodes together with a
three-dimensional model of the entire device structure. (b) Energy level
diagram of the double quantum dot for positive bias. (c) Measurement of the few-electron double
quantum dot stability diagram at a finite bias of +5mV and for weak interdot
coupling. The baseline of the  lower left triangle indicates alignment between
the two ground state levels in the quantum dots G(1,0) and G(0,1). Red arrows
indicate a change in detuning $\Delta$ and overall energy $\varepsilon$ as a
function of the two plunger gate voltages $V_{\rm lg2}$ and  $V_{\rm
lg4}$. (d) Close-up of the resonance edge between G(0,1) and G(1,0) for strong
coupling ($V_{\rm lg3}=-0.55$V). The upper-left inset shows the measured
current $I_{\rm SD}$, while its derivative with respect to  $V_{\rm lg4}$ is
displayed in the main panel to highlight the resonance lines running parallel
to the ground state transition line.}
\end{figure}

In order to probe the individual phonon modes, we apply the resonance condition
of tunneling between two quantum dots (QDs) inside a 
nanowire \cite{FasthNL2005,FuhrerNL2007}.
Transport properties of gate-defined lateral double QDs in
two-dimensional electron gases have been widely studied (see
Ref.~\onlinecite{WielRMP2003} and references cited therein). The main feature
is the strong enhancement of transport if two QD levels are aligned by tuning
the corresponding plunger gate voltages. 
The role of phonon scattering 
in these lateral QD systems has been investigated 
experimentally
\cite{VanDerVaartPRL1995,FujisawaScience1998} as well as theoretically
\cite{BrandesPRL1999PR2005}, revealing interesting effects such as
oscillations due to the phonon interference of the spatially separated QDs. 
While these studies essentially probed the bulk phonon structure of the
substrate, we show that the confined phonon modes in the investigated nanowire appear 
as satellite peaks to the resonance for tunneling between the QDs.


The experimental setup is shown in Fig.~\ref{FigExpStruct}(a): A
catalytically grown InAs nanowire with a diameter of 50 nm is placed on a
substrate with a pattern of metallic gates (periodicity of 60 nm), covered by
an 18 nm SiN film. The sample is identical to the one in
Ref.~\onlinecite{FasthPRL2007}, where the focus was on single QD
properties. Here, the first, third, and fifth gate (shown in red color) define
an electrostatic confinement, leading to two spatially separated QDs in the
wire. The other two gates (shown in green color) are used as plungers to tune
the QD filling. A bias voltage $V_{\rm bias}=\pm5$~mV is applied across the
structure [see Fig.~\ref{FigExpStruct}(b)]. The measurements are performed at
60~mK in a dilution refrigerator. The strong confinement in both the radial
and growth direction allows for a control of the number of electrons in each
QD down to the last one \cite{FasthPRL2007,epaps}. Figure~\ref{FigExpStruct}(c)
shows that varying the potential of the two plunger gates generates the
typical stability plot for double QD systems \cite{WielRMP2003}.

In Fig.~\ref{FigExpStruct}(d), we focus on the 
transition region between the states G(0,1) and G(1,0), where the energetically lowest
level of one of the QDs is occupied, while Coulomb charging hinders the
addition of a second electron to either QD.  The bright line, constituting the
common base of the double triangle, indicates the relation between the plunger
gate voltages where the levels G(1,0) and G(0,1) of the two QDs align,
thus enhancing resonant tunneling through the double QD as long as the levels
are located between the electrochemical potentials $\mu_{L/R}$ in the
left/right part of the nanowire,  respectively, see
Fig.~\ref{FigExpStruct}(b). Below this line, there is also substantial current
flow which can be attributed to inelastic processes
\cite{VanDerVaartPRL1995}. Our key observation is the presence of  pronounced
lines running parallel to the ground state resonance line in this region. On
these lines, the QD levels maintain a fixed detuning $\Delta$ from resonance
with an average spacing of $\delta \approx 180-200$~$\mu$eV. These parallel
lines are found for both bias directions and cannot be attributed to excited
states which are several meV above the ground state levels \cite{epaps}.
In conventional double QD
systems based on a two-dimensional electron gas, similar features have been
found  \cite{FujisawaScience1998} and attributed to an interference in the
electron-phonon interaction \cite{BrandesPRL1999PR2005} with the bulk phonons.
As discussed below, our findings cannot be explained by this
interference. Instead they result from the particular phonon structure in
nanowires. Thus, the observation of these current replicas manifests the
interplay between electron and phonon confinement in the system.

 
To model the phonon scattering environment in the nanowire, we calculate
the one-dimensional phonon modes of the nanowire within an isotropic elastic
continuum model \cite{StroscioJAP1994Book2001}. Our InAs nanowires 
typically have nearly perfect wurtzite crystal structure with the wire
oriented along the $\langle0001\rangle$ direction. Neighboring stacking layers
of the wurtzite structure agree with those in a zinc-blende structure in
$\langle111\rangle$ direction \cite{BykhovskiAPL1996}. Since material
parameters for the InAs wurtzite lattice are largely missing, we take them
from a rotated bulk zinc-blende structure instead. Calculations of strain in
nanowires have shown that such a transformation between the two lattices is a
good approximation \cite{LarssonNT2007}. The only material parameters which
enter the calculation of the phonon modes are the  mass density
$\rho_m = 5680$~kg/m$^3$ and the  longitudinal ($v_l = 4410$~m/s) and
transverse ($v_t = 2130$~m/s) sound velocity, which are taken along 
$\langle111\rangle$. Free-surface
boundary conditions are assumed so that the strain vanishes at the surface of
the wire. We neglect the influence of the substrate and the Ohmic contacts on the phonon mode
structure.

\begin{figure}
\includegraphics[width=0.8\linewidth]{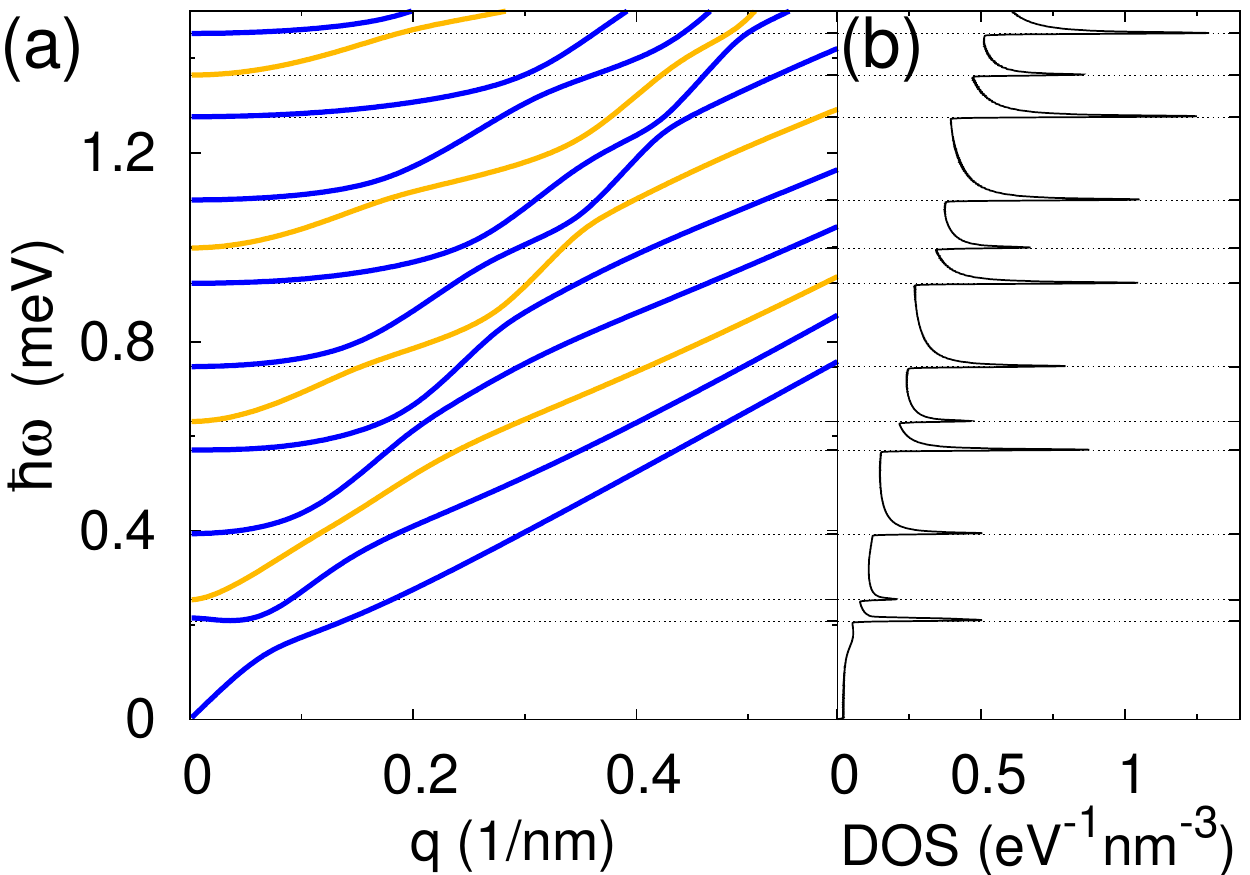}
\caption{\label{FigPhonDispDOS} (Color online) (a) Phonon spectrum of an InAs nanowire with
  radius $R = 25$~nm (compressional modes only). The blue/orange lines depict
  modes which are of axial/radial character for $q\to0$.
  (b) Corresponding density of states of the phonon spectrum.}
\end{figure}

Figure~\ref{FigPhonDispDOS}(a) shows the compressional (dilatational) phonon
modes \cite{FootnoteModes} for our nanowire structure as a function of the
one-dimensional phonon quasi-momentum $q$ resulting from the translation
invariance along the nanowire axis. These modes decouple into purely radial
and purely axial modes for $q \to 0$. The corresponding density of states
(DOS) is shown in Fig.~\ref{FigPhonDispDOS}(b). The strong peaks in the DOS
are roughly equidistant with an energy separation of about $180$~$\mu$eV,
fitting the replicas in the current measurement well. It should be noted that
the singularities in the DOS are a unique feature of the one-dimensionality of
the nanowire.

\begin{figure}
\includegraphics[width=\linewidth]{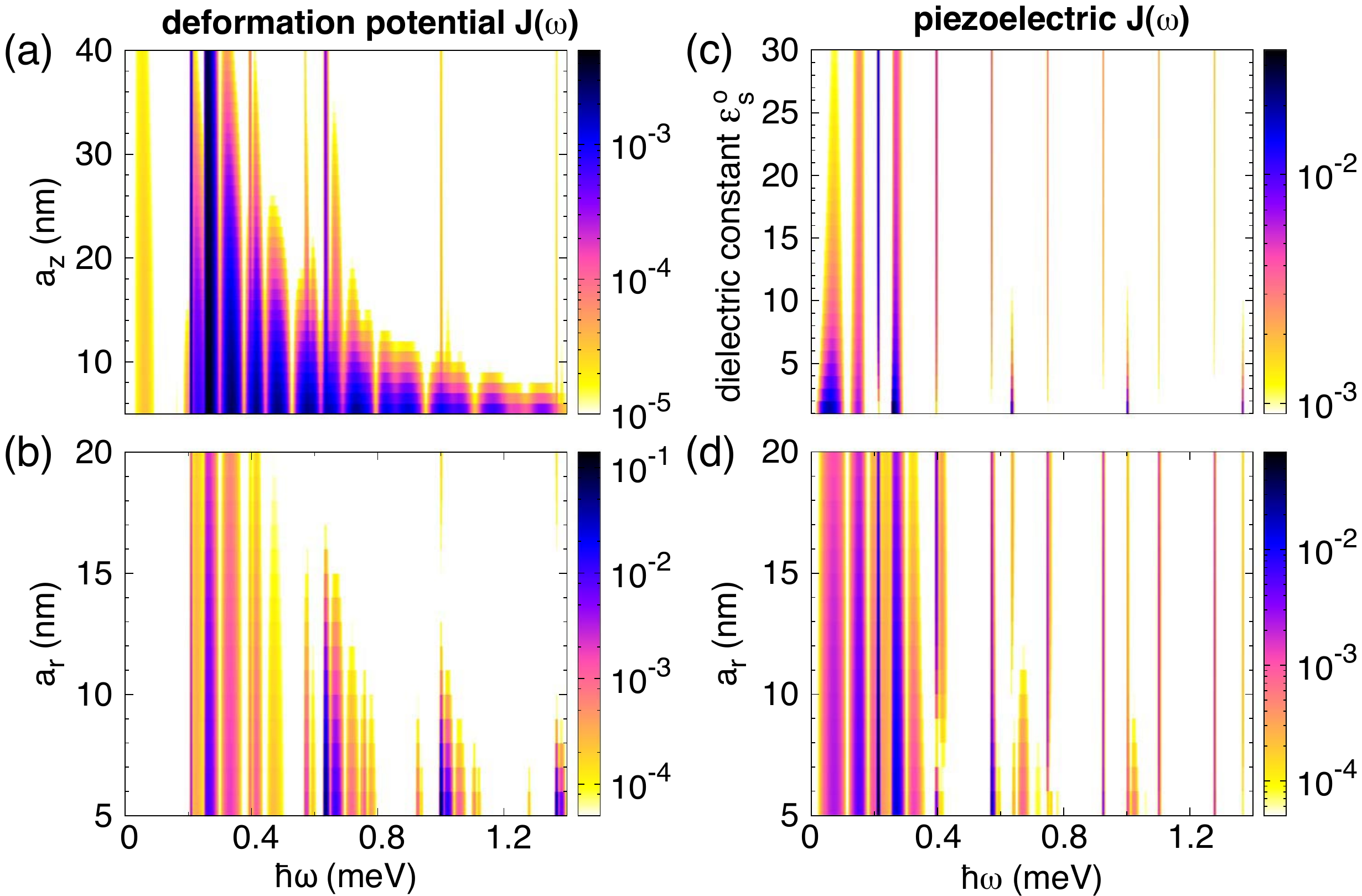}
\caption{\label{FigPhonCoup} (Color online) Phonon spectral density $J(\omega)$ (color scale
  in meV)
  for different
  scattering mechanisms and parameters: Deformation potential coupling for
  varying (a) axial confinement length $a_z$ and (b) radial confinement  
  length $a_r$. Piezoelectric coupling for varying
  (c) dielectric constant $\eps_s^o$ outside the wire and (d)
  radial confinement length $a_r$. The 
  corresponding other parameters are fixed to the values discussed in the
  text.}
\end{figure}

Now we focus on the coupling strength between the 
different phonon modes and the electronic states in the QDs.
This coupling strength can be quantified by the
phonon spectral density \cite{LeggettRMP1987}
\begin{equation}
J(\omega) = \frac{1}{\hbar^2} \sum_{q \kappa} 
|M^{q \kappa}_{1} - M^{q \kappa}_{2}|^2 \delta(\omega - \omega_{q \kappa})\, .
\label{EqJomega}
\end{equation}
$M^{q \kappa}_i$ describes the electron-phonon coupling element of the
electronic state in dot $i$ to the phonon mode $\kappa$ with one-dimensional
quasi-momentum $q$; here, we assume diagonal electron-phonon coupling.
$J(\omega)$
takes into account the effective coupling strength $|M^{q \kappa}_{1} - M^{q
\kappa}_{2}|$ between the two QDs as well as the phonon DOS addressed above. 
We consider  deformation potential and piezoelectric coupling to the phonons, 
for details see Ref.~\onlinecite{WeberPSSB2009}. The electronic states of
the QDs are modeled by  cylindrical Gaussians with radial and axial
confinement lengths $a_r$ and $a_z$, respectively, which is appropriate for a
harmonic confinement. From the charging spectrum as well as previous measurements \cite{FasthPRL2007}, we choose the values of $a_r = 15.0$~nm and $a_z
= 20.0$~nm. Again we use the parameters from the InAs zinc-blende structure:
conduction band deformation  potential $D_c = -5.08$~eV, static dielectric
constant $\eps_s^i = 15.15$, and piezoelectric constant
$e_{14}=-0.115$~C/m$^2$ (Ref.~\onlinecite{BesterPRL2006}). Note that the
latter value is uncertain as a variety of different values are given in the
literature. Taking into account the orientation of the nanowire, this provides
the wurtzite values $e_{15} = e_{31} = -e_{33}/2 = -e_{14}/\sqrt{3}$
\cite{BykhovskiAPL1996}. For the calculation of the piezoelectric coupling,
the boundary condition for the piezoelectric field at the surface of the wire
is of crucial importance. The wires are located on an 18 nm thick layer of
SiN. Below this layer, metallic contacts are present, corresponding to
$\eps_s\rightarrow \infty$. In our radially symmetric model
\cite{WeberPSSB2009}, we approximate these features by assuming an outer
dielectric material with $\eps_s^o= 20$. However, different values are also
studied below. We assume zero temperature in analogy to the experimental
situation. Fig.~\ref{FigPhonCoup} displays the numerical results 
for $J(\omega)$, see Eq.~(\ref{EqJomega}), for different
QD sizes and varying values of $\eps_s^o$.

The deformation potential [Fig.~\ref{FigPhonCoup}(a,b)] couples mainly to the
radial modes via the diagonal elements of the strain tensor
\cite{WeberPSSB2009}. Thus only some of the peaks from
Fig.~\ref{FigPhonDispDOS} lead to pronounced electron-phonon coupling, and the
experimentally observed peak spacing cannot be explained by restricting to
this scattering mechanism. Reducing the QD sizes $a_r,a_z$, these peaks
receive tails at higher energies as the matrix elements for larger $q$, i.e.,
higher $\omega$, increase with the reduction in QD size.  Due to interference
between both QDs, the matrix element $|M^{q \kappa}_{1} - M^{q \kappa}_{2}|$
vanishes if $qd/2\pi$ is an integer, where $d=120$~nm is the separation
between the QDs. This causes pronounced oscillations in the spectral density
$J(\omega)$, which have been studied before for bulk phonons \cite{FujisawaScience1998,BrandesPRL1999PR2005}. The corresponding
energy oscillation period depends on the velocity $v$ of the phonon mode and
takes values between $\delta = \hbar v2\pi/d\approx 120$~$\mu$eV for the
acoustic mode for $q \rightarrow 0$ and $\delta \approx 70$~$\mu$eV  for the
transverse sound velocity. Even smaller velocities are possible close to
extrema in the dispersion relation. Thus, these energies are too small to
explain the origin of the experimentally observed current oscillations in
Fig.~\ref{FigExpStruct}(d).

The piezoelectric coupling also couples dominantly to the radial modes (mainly
via the coefficient $e_{31}$) for a small outer dielectric constant
$\eps_s^o$ [Fig.~\ref{FigPhonCoup}(c)]. With increasing $\eps_s^o$, the axial
modes (main coupling via $e_{15}$) provide distinct peaks at their dispersion
extrema, which dominate the coupling if $\eps_s^o$ becomes of the order of
the
dielectric constant inside the wire $\eps_s^i$.  In the latter case, we
recover a spacing of $\delta \approx 180 $~$\mu$eV, in good agreement with the
experimental data. A special situation arises for mode 1, which is axial but
mainly couples via $e_{33}$ since its displacement is approximately
constant. Here, increasing $\eps_s^o$ causes a relative shift of coupling strength to
larger energies. For increasing radial confinement length
$a_r$, the coupling strength typically drops for both the deformation
potential and piezoelectric coupling [Fig.~\ref{FigPhonCoup}(b,d)]. However,
the dependence of the piezoelectric coupling is much weaker than that of the
deformation potential.

\begin{figure}
\includegraphics[width=0.8\linewidth]{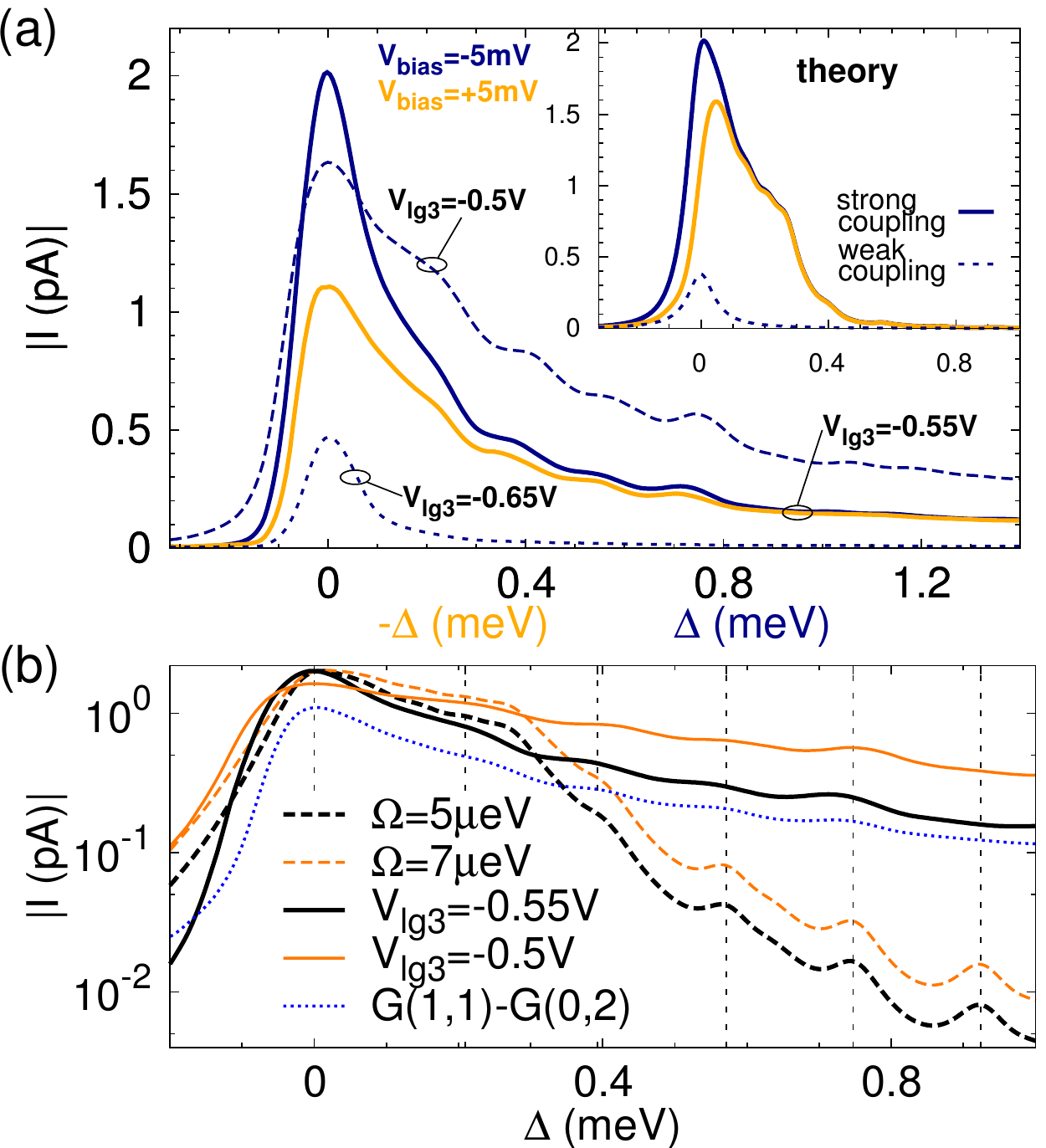}
\caption{\label{FigCurr} (Color online) Current as a function of the detuning
  between the two QDs. (a) Experimental data: In the weak
  coupling case ($V_{\rm lg3}=-0.65$~V, dotted line) a single nearly
  symmetric resonance occurs. For strong coupling [$V_{\rm lg3}=-0.55$~V,
  solid lines, data as Fig.~\ref{FigExpStruct}(d), and  $V_{\rm lg3}=-0.5$~V,
  dashed line] a right tail appears showing
  oscillatory behavior. For $V_{\rm lg3}=-0.55$~V both bias directions are
  shown by different colors. The inset shows numerical results for our main
  parameters (solid lines) and for weak coupling $\Omega = 200$~neV,
  $\Gamma_L=\Gamma_R=100$~$\mu$eV (dotted line). (b) Comparison of 
  peak positions indicated by vertical lines: 
  Experimental data (solid lines) and theory (dashed lines) for
  different interdot couplings for the G(0,1)-G(1,0) resonance. The dotted
  line shows data for the  G(0,2)-G(1,1) resonance.}
\end{figure}

Figure~\ref{FigCurr}(a) shows the experimental current as a function of
the detuning $\Delta$ between the two QD levels for the G(0,1)-G(1,0) 
resonance. Roughly equidistant
shoulders/peaks (with a separation $\delta \approx 180-200$~$\mu$eV) are clearly
observed in the tail to the right of the main peak for both bias directions and strong interdot coupling ($V_{\rm lg3}=-0.55$~V and $-0.5$~V). Upon increasing the  barrier
height ($V_{\rm lg3}=-0.65$~V), the current peak becomes more symmetric and no
side peaks are observed.

In our theoretical model, we calculate the current through the double QD
system following Ref.~\onlinecite{BrandesPRL1999PR2005}. The (diagonal)
electron-phonon interaction is treated within the independent Boson model
\cite{MahanBook1990}, and the current is calculated for a (weak) coupling to
the left and right contacts, $\Gamma_{L/R}$, using a non-perturbative
treatment of the interdot coupling $\Omega$. A finite phonon lifetime
$\gamma_{\rm ph} = 40$~$\mu$eV is taken into account following the work of
Refs.~\onlinecite{ZimmermannICPS2002ForstnerPSSB2002}. In the following, we
assume $\Omega = 5$~$\mu$eV, $\Gamma_R =10$~neV, and $\Gamma_L = 90$~neV,
which fits the measured peak current and provides a part of the asymmetry
between forward and reverse bias (as the model neglects spin, the full
asymmetry cannot be accounted for). The results are shown in the inset of
Fig.~\ref{FigCurr}(a), solid lines, where the main peak  at $\Delta\approx 0$
has a pronounced tail at positive $\Delta$ due to tunneling combined with
phonon emission. The shoulder until $\Delta\approx 0.25$ meV combines the
acoustic phonon mode (extending to zero frequency) with its first interference
oscillation peak, the axial mode 2, and the radial mode 3 (these
different features are just about discriminable in the theory but appear
washed out in the experiment). For larger detunings, the peaks can be
attributed to the onset of the higher axial modes with an energy separation of
$\delta \approx 180$~$\mu$eV, in agreement with the experiment. These features
disappear in the weak coupling case with $\Omega\ll \Gamma$ (dotted line). In
Fig.~\ref{FigCurr}(b), we show that the peak positions agree between
experiment and simulation for different couplings
\cite{FootnoteCurrent}. The same peak positions are also found at other
resonances in this double QD [see the dotted line in
Fig.~\ref{FigCurr}(b) and further data from the same sample presented in Ref.~\onlinecite{epaps}].
Thus, these peak positions do not depend on the particular
electronic states involved. This strongly supports our interpretation that the
observed peak structure is a signature of the phonon confinement.


In {\em conclusion}, we have found a strong interplay between the confined
electronic and phononic systems in catalytically grown nanowires. Our current
spectroscopy in the few-electron regime of a nanowire-based double
quantum dot allows us to probe the confined phonon environment of the
nanowire. The characteristic peak structure can be well explained
theoretically if the piezoelectric coupling mechanism is taken into account.

\begin{acknowledgments}
We acknowledge financial support by the Swedish Research Council (VR), the Swedish Foundation for Strategic Research (SSF), and the Knut and Alice Wallenberg Foundation (KAW), and
thank O. Karlstr{\"o}m, M. Bj{\"o}rk, D. Loss, and B. Coish for valuable discussions.
\end{acknowledgments}


\begin{thebibliography}{10}
\bibitem{FootnoteModes}
The torsional modes are azimuthally directed and do not couple to the
  electronic system in the case of the considered lattices. The flexural modes
  exhibit an azimuthal dependence and thus do not couple to the electronic QD
  ground states, which are azimuthally symmetric.
\end{thebibliography}

\end{document}